\documentclass[letterpaper]{article} 
\usepackage{aaai2026}  
\usepackage{times}  
\usepackage{helvet}  
\usepackage{courier}  
\usepackage[hyphens]{url}  
\usepackage{graphicx} 
\usepackage{natbib}  
\usepackage{caption} 
\usepackage{algorithm}
\usepackage{algorithmic}

\usepackage{newfloat}
\usepackage{listings}
\DeclareCaptionStyle{ruled}{labelfont=normalfont,labelsep=colon,strut=off} 
\floatstyle{ruled}
\newfloat{listing}{tb}{lst}{}
\floatname{listing}{Listing}
\nocopyright

\usepackage{latexsym}
\usepackage{amssymb}
\usepackage{amsmath}
\usepackage{amsthm}
\usepackage{enumitem}
\usepackage{color}
\usepackage{diagbox}
\usepackage{afterpage}

\usepackage{booktabs}
\usepackage{multirow}
\newcommand{\minisection}[1]{\vspace{3pt}\noindent\textbf{#1.}}

\title{Synthesize, Retrieve, and Propagate: A Unified Predictive \\ Modeling Framework for Relational Databases}
\author {
    Ning Li\textsuperscript{\rm 1}\thanks{Work was done during an internship at AWS Shanghai AI Lab.},
    Kounianhua Du\textsuperscript{\rm 1},
    Han Zhang\textsuperscript{\rm 1}, \\
    Quan Gan\textsuperscript{\rm 2}, 
    Minjie Wang\textsuperscript{\rm 2},
    David Wipf\textsuperscript{\rm 2},
    Weinan Zhang\textsuperscript{\rm 1}
}
\affiliations {
    \textsuperscript{\rm 1}Shanghai Jiao Tong University \quad
    \textsuperscript{\rm 2}AWS Shanghai AI Lab\\
    lining01@sjtu.edu.cn
}

\usepackage{bibentry}

\begin{document}

\maketitle

\begin{abstract}
Relational databases (RDBs) have become the industry standard for storing massive and heterogeneous data. However, despite the widespread use of RDBs across various fields, the inherent structure of relational databases hinders their ability to benefit from flourishing deep learning methods.
Previous research has primarily focused on exploiting the unary dependency among multiple tables in a relational database using the primary key - foreign key relationships, either joining multiple tables into a single table or constructing a graph among them, which leaves the implicit composite relations among different tables and a substantial potential of improvement for predictive modeling unexplored.
In this paper, we propose SRP, a unified predictive modeling framework that \textbf{S}ynthesizes features using the unary dependency, \textbf{R}etrieves related information to capture the composite dependency, and \textbf{P}ropagates messages across a constructed graph to learn adjacent patterns for prediction on relation databases. By introducing a new retrieval mechanism into RDB, SRP is designed to fully capture both the unary and the composite dependencies within a relational database, thereby enhancing the receptive field of tabular data prediction.
In addition, we conduct a comprehensive analysis on the components of SRP, offering a nuanced understanding of model behaviors and practical guidelines for future applications. Extensive experiments on five real-world datasets demonstrate the effectiveness of SRP and its potential applicability in industrial scenarios. The code is released at https://github.com/NingLi670/SRP.
\end{abstract}


\section{Introduction}
The advent of the digital age has led to an exponential increase in the generation and storage of data, which is predominantly housed in relational databases due to their efficiency and structured design \citep{zahradnik2023deep}. Relational databases (RDBs) \citep{chamberlin1976relational} have become the industry standard across diverse fields such as recommendation systems \citep{sarwat2013recdb}, engineering \citep{yaqoob2022blockchain}, and enterprise applications \citep{halpin2010information}. 

Despite the widespread use of RDBs, the application of advanced machine learning models, particularly neural models, on this form of data is limited \cite{zhang2023gfs} due to the inherent structure. Unlike the fixed-size numeric tensors used in machine learning models, an RDB consists of interlinked table structures with heterogeneous features. This discrepancy necessitates different approaches to processing RDB data and learning its representation. 

\begin{figure}[t]
    \centering
    \includegraphics[width=1\linewidth]{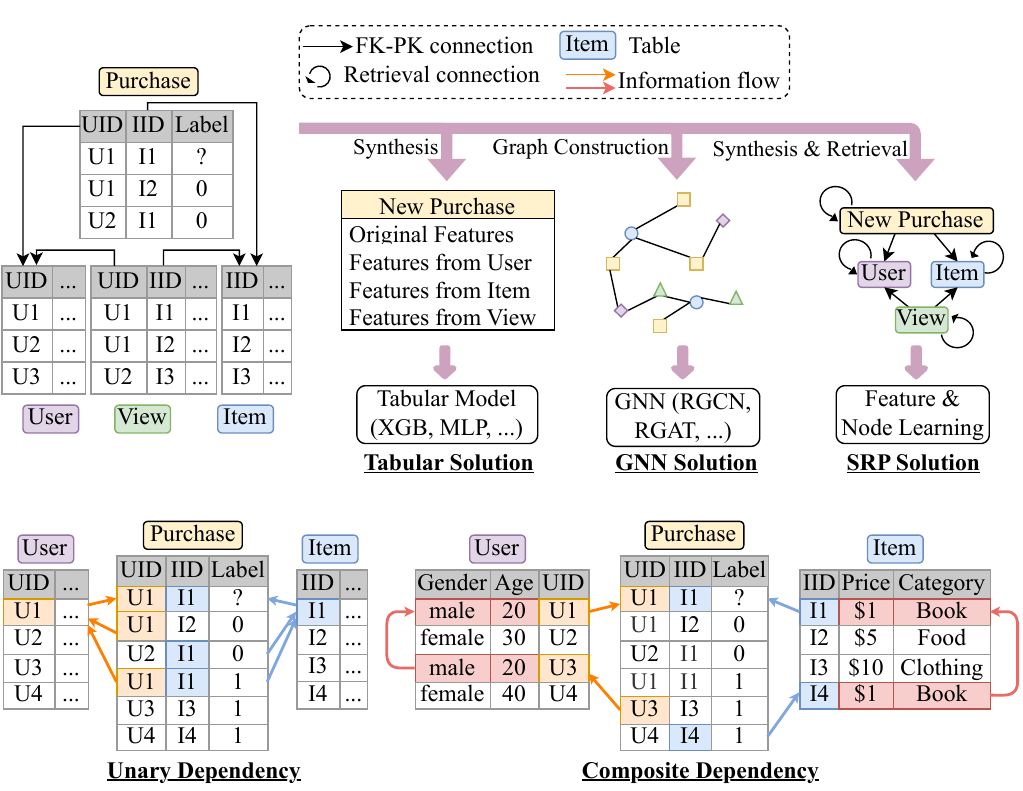}
    \caption{Illustration of different solutions and difference between unary and composite dependencies.} 
    \label{fig:motivation}
\end{figure}

Previous works extract the dependencies among tables within an RDB using foreign key - primary key (FK-PK) relationships, either by merging multiple tables into a single table or constructing a graph from multiple tables, followed by a traditional tabular model or graph neural network for predictions, as illustrated in the upper part of Figure \ref{fig:motivation}. This dependency pattern is unary, constrained by the inherent connections between the unary FK column and PK column in an RDB. As depicted in the lower left of Figure \ref{fig:motivation}, the target instance with an unknown label can only gather information from instances sharing the same user or item along the inherent FK-PK connections. However, the implicit composite relations within an RDB still remain unexplored, which could be of great help for prediction. As shown in the lower right of Figure \ref{fig:motivation}, by retrieving similar instances using composite attributes, the target instance can access information that is not explicitly connected in the original RDB. Motivated by this, we aim to design a framework that extensively explores these richer dependencies within an RDB to achieve better prediction performance.

In this paper, we propose Synthesis-Retrieval-Propagation (SRP), a novel unified predictive framework that captures both the unary and the composite dependencies among relational databases, enlarging the receptive field of tabular data prediction. Conceptually, \textit{unary dependency} refers to the direct relationship between entities through FK-PKs, typically constructed as the connections between two unary columns, whereas \textit{composite dependency} refers to the complex, often indirect relationship between entities that are not explicitly defined by FK-PK connections in the database schema. These dependencies arise from composite attributes and contextual similarities between entities, which can be discovered through methods such as retrieval-based similarity measures \cite{qin2021retrieval, zheng2023dense}.

The SRP framework consists of three key modules: synthesis, retrieval, and propagation. 
Given a target instance, i.e., a particular row in the target table of the considered RDB, the synthesis module generates additional features along the FK-PK connections from related tables in the RDB, extracting unary dependencies. The retrieval module builds new relationships beyond the original FK-PKs to capture the implicit composite dependencies by performing retrieval and top-$K$ ranking based on the rich attributes of entities. The propagation module then learns adjacent patterns, utilizing the retrieval information through message passing. The final prediction is made by integrating the representations from different modules, effectively fusing the unary and composite dependencies. 
As a result, SRP yields a solution model that traverses all tables within the  RDB, synthesizing new features and constructing new connections beyond the existing FK-PK relationships. This enables SRP to capture both explicit and implicit information from all tables, constructing a comprehensive representation of the data.

The main contributions can be summarized in three folds:
\begin{itemize}
\item \textbf{SRP framework}. We propose a framework named SRP that captures both the unary and the composite dependencies within a relational database through synthesis, retrieval, and propagation. This unified framework is adaptable to various types of RDBs.
\item \textbf{Introduce retrieval to RDB prediction}. Leveraging the retrieval algorithm, we uncover the implicit connected information beyond the inherent FK-PK relationships for modeling complex composite dependencies in RDBs. To the best of our knowledge, SRP is the first framework to introduce the retrieval mechanism for predictive tasks in relational databases. 
\item \textbf{SOTA performance and comprehensive analysis}. Experiments on five real-world datasets demonstrate that SRP consistently outperforms a wide range of competitive baselines, achieving state-of-the-art performance. Furthermore, our empirical analysis reveals that RDBs with diverse characteristics, such as limited attributes, sparse relational structures, or imbalanced distributions, exhibit different preferences for the SRP modules.
\end{itemize}

\section{Related Works} \label{sec:related-work}
\minisection{Feature Engineering}
Early approaches relied on manual feature engineering for predictive tasks on relational databases~\cite{chen2002data, nisbet2009handbook}, but this process is labor-intensive and heavily dependent on domain expertise. To automate this process, methods such as DFS~\cite{kanter2015deep}, OneBM~\cite{lam2017one}, R2N~\cite{lam2018neural}, and DAFEE~\cite{zhao2020dafee} have been proposed. These techniques construct features by following join paths or table relationships, using either rule-based transformations or neural models. However, they primarily capture unary dependencies along FK-PK paths and often rely on simple aggregation strategies, which may discard important information. In contrast, our SRP framework introduces a retrieval mechanism to build richer, cross-table dependencies and employs a frequency-aware aggregator to better preserve valuable patterns.

\minisection{Relational Deep Learning}
Recent works have focused on converting relational databases into graphs. Methods like RDB2Graph and 4dbinfer~\cite{wang4dbinfer} represent table rows as nodes or edges, enabling the use of heterogeneous GNNs such as RGCN~\cite{schlichtkrull2018modeling}, RGAT~\cite{busbridge2019relational}, HGT~\cite{hu2020heterogeneous}, PNA~\cite{corso2020principal}, and GFS~\cite{zhang2023gfs} for prediction tasks. Other works, including ATJ-Net~\cite{bai2021atj}, Blueprint~\cite{zahradnik2023deep}, and SPARE~\cite{fey2024relational}, construct hypergraphs over RDBs to capture higher-order dependencies through deep graph learning. Benchmark efforts such as 4DBInfer~\cite{wang4dbinfer} and RelBench~\cite{fey2024relational} have further encouraged systematic evaluation in this domain. In SRP, we also include graph neural networks in our solution space. Unlike previous works, which only construct graphs based on original FK-PK connections, we cooperate GNN with a retrieval mechanism to expand the receptive fields. With the synthesis, retrieval, and propagation module, SRP can capture both the unary and composite dependencies among relational databases.

While several studies have explored neural approaches for relational data imputation~\cite{spinelli2020missing} and query answering~\cite{hilprecht2019deepdb}, these works target different objectives. Our focus is supervised predictive modeling over relational databases, rather than data completion or fact retrieval.

\begin{figure*}
\centering 
\includegraphics[width=1\linewidth]{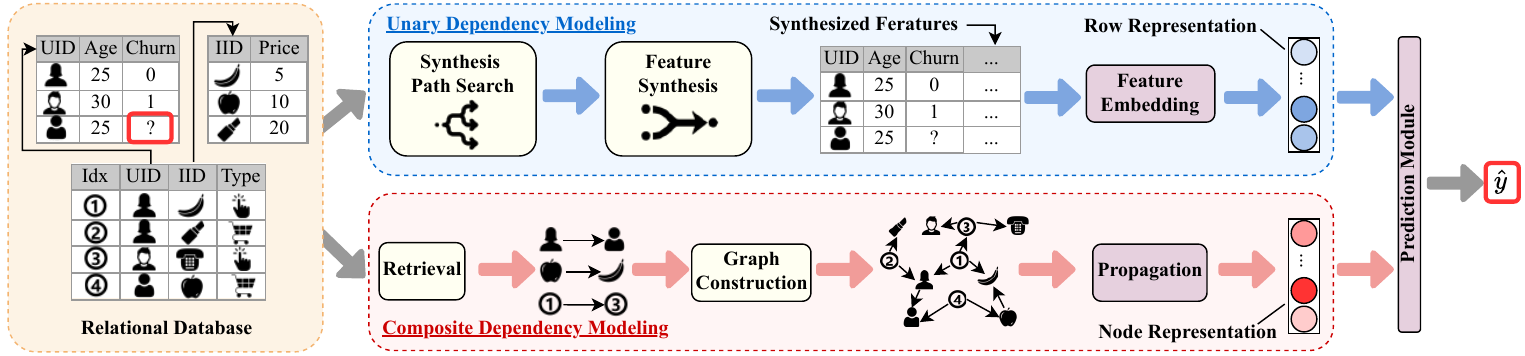} 
\caption{The SRP framework. An RDB is fed into the unary dependency (Blue) and composite dependency (Red) modeling parts in parallel to generate corresponding representations, then the final prediction $\hat{y}$ is generated by the prediction module.  The yellow blocks represent the offline process, while the purple blocks are updated during model training.}
\label{fig:framework}
\end{figure*}

\minisection{Retrieval Techniques on Tabular Data}
Retrieval-based models have been widely used in tabular data tasks. SIM~\cite{pi2020search} and RIM~\cite{qin2021retrieval} retrieve similar samples to model user interests and cross-row/column patterns, while DERT~\cite{zheng2023dense}, TABR~\cite{gorishniy2024tabr}, and PET~\cite{du2022learning} enhance representation learning via dense retrieval or hypergraph propagation. However, these methods are limited to single-table settings. In contrast, our SRP framework is the first to apply retrieval augmentation in multi-table relational databases.


\section{SRP Framework}

\subsection{Task Formulation}
A relational database (RDB) is a collection of tables $D = \{T^1, T^2, \dots, T^N\}$. Let $T^n_{i:}$ and $T^n_{:j}$ represent the $i$-th row (entity) and the $j$-th column (attribute) of the $n$-th table $T^n$, respectively. Thus, $T^n_{i,j}$ represents the value of row $i$ and column $j$ in table $T^n$. In an RDB, different tables are interconnected through primary keys (PKs) and foreign keys (FKs). A column $T^n_{:i}$ containing unique values in each row can serve as a PK of table $T^n$, while a foreign key $T^n_{:j}$ is a column whose values correspond to the PK values of another table. 

The prediction task over an RDB is to predict a target variable $T^t_{i,j}$ of the target table $T^t \in D$ given the relational database $D$ and all relationships between tables. Usually, all values in the target column $T^t_{:,target}$ need to be predicted.

\subsection{Framework Overview}
As shown in Figure \ref{fig:framework}, SRP comprises Synthesis, Retrieval, and Propagation modules, forming two parallel workflows for modeling unary and composite dependencies. Given an RDB, the synthesis module traverses all tables following FK-PK connections to collect attributes from other tables that can be integrated into the target table. Since the tables have different sets of columns, it is non-trivial to augment the target table with other tables; hence, the synthesis module employs operations like join and aggregation to achieve feature synthesis. The synthesized features are then embedded to make final predictions. 

In the composite dependency modeling workflow, the retrieval module establishes new connections by finding a fixed number of similar rows for each row based on a similarity metric. By digging new relationships in addition to FK-PK pairs, the retrieval module is able to uncover composite dependencies.
After retrieval, such relationships together with the PK-FKs are leveraged to convert the RDB into a heterogeneous graph, followed by the propagation module to perform message passing and produce a node embedding for each target sample.  Finally, the node and feature embeddings are combined by the prediction module to fuse unary and composite dependencies for final predictions.

\subsection{Unary Dependency Modeling}\label{sec:unary-dependency-modeling}

The synthesis module focuses on unary dependency modeling, which can be separated into synthesis path search, feature synthesis, and feature embedding.

\begin{figure}[htbp]
\centering 
\includegraphics[width=1\linewidth]{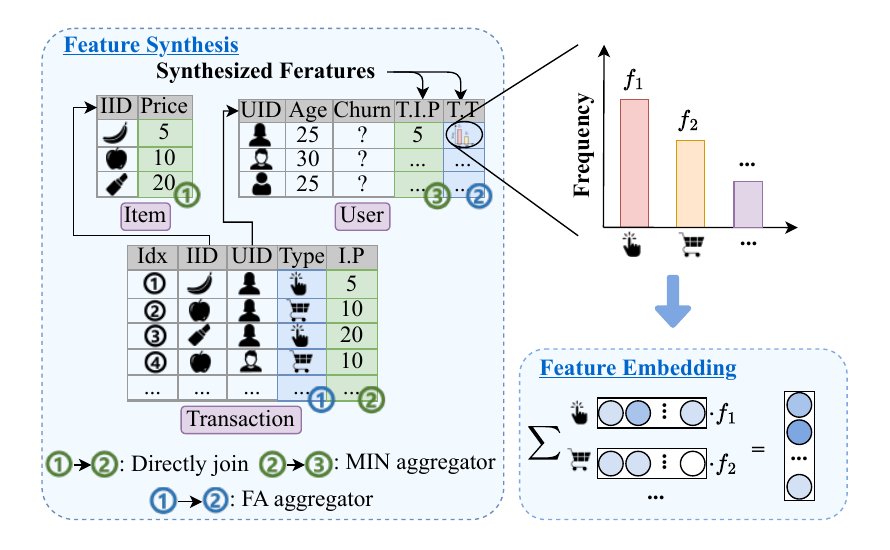} 
\caption{Illustration of the synthesis module.}
\label{fig:synthesis}
\end{figure}

\subsubsection{Deep Feature Synthesis}
We use Deep Feature Synthesis \cite{kanter2015deep} and our own extensions to complete synthesis path search and feature synthesis. Starting from the target table, it traverses the tables in a depth-first search manner following the FK-PK connections between tables to find synthesis paths. Features from other tables will be joined or aggregated to the target table along the paths. In Figure \ref{fig:synthesis}, the left shows how \texttt{Price} and \texttt{Type} columns are synthesized to the target \texttt{User} table. When aggregating values from other tables, we select the following aggregators: \verb|COUNT| generates the number of elements per aggregation. \verb|MEAN|, \verb|MAX|, and \verb|MIN| aggregators are used for numeric features to capture the central tendency and range. Text features are combined using simple \verb|JOIN| to preserve the full content. For categorical features, \verb|MODE| is used to record the most frequent category. It is worth noting that during the aggregation process for each target sample, no future information from other tables is incorporated, thereby preventing any information leakage.

In tabular representation learning, categorical features usually carry significant discriminative information for prediction. However, the \verb|MODE| aggregator only retains the most common category, ignoring the presence of other potentially relevant categories. Here we introduce a frequency-aware aggregator \verb|FA| to record all categorical values and their occurrences, as is shown in Figure \ref{fig:synthesis} right. Cooperating with downstream feature encoders, it aims to restore the context about categorical distribution as effectively as possible. We will present the details in Section \ref{sec:feature-embedding}.

\subsubsection{Feature Embedding}\label{sec:feature-embedding}
 After feature synthesis, the new features and the original features in the target table will then be encoded as feature embeddings. We use linear projection for numeric features, one-hot encoding for categorical features, and \verb|glove| \cite{pennington2014glove} to transform text into vector embeddings. For the features generated by the \verb|FA| aggregator, assume that there are $n$ categorical values $\{c_1, c_2, \dots, c_n\}$ and their occurrences $\{a_1, a_2, \dots, a_n\}$. During the encoding phase, we keep the $m$ most frequently occurring categories $\{c_{(1)}, c_{(2)}, \dots, c_{(m)}\}$ and normalize their occurrences into frequencies $f$ as the weight to aggregate the one-hot encoding of each category. Formally, we have 
\begin{equation}
    \text{Encoded Feature}=\sum_{i=1}^m(\frac{a_{(i)}}{\sum_{j=1}^ma_{(j)}}\cdot \text{one-hot}(c_{(i)})),
\end{equation}

The overall process of the synthesis module is represented as 
\begin{equation}
    \mathbf{H_u} = E_u(S(D), T^t;\theta),
\end{equation}
where $S$ is the synthesis module, $S(D)$ and $T^t$ represents the synthesized and original features, $\theta$ is the parameters. $E_u$ is the concatenation of a series of encoders $E^1_u\oplus E^2_u\oplus\dots\oplus E^{N_s}_u$ corresponding to each feature, where $N_s$ is the total number of features and $\oplus$ means concatenation.

\subsection{Composite Dependency Modeling}
The composite dependency modeling workflow focuses on capturing complex relationships within an RDB that go beyond the explicit FK-PK relationships. Here we introduce the retrieval and propagation modules, which are designed to enhance node representation by incorporating information from retrieved similar instances and adjacent neighbors.

\subsubsection{Retrieval}\label{sec:retrieval}

\begin{figure}[htbp]
\centering 
\includegraphics[width=1\linewidth]{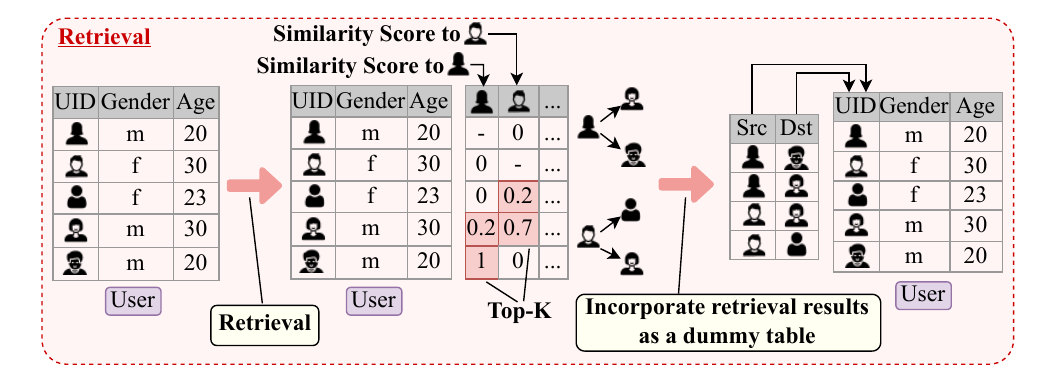} 
\caption{The Retrieval process for each table in RDB.}
\label{fig:retrieval}
\end{figure}

The retrieval module aims to capture composite dependencies in relational databases by retrieving similar rows within each table and establishing new connections accordingly. When performing retrieval on a table, each row will be retrieved to find similar rows. We call the retrieved row $T^X_{q:}$ as the query row, and all other rows $T^X_{\setminus q:}$ are keys, as well as values. For simplicity, we represent $T^X_{q:}=\{x^q_1, x^q_2, \dots, x^q_F\}$, where $x^q_i$ is the $i$-th value of row $q$. Following \cite{qin2021retrieval, du2022learning}, we use BM25 \cite{robertson1995okapi} algorithm to calculate the similarity and retrieve top-$K$ similar rows. Specifically, for categorical features, the similarity score between query $x^q$ and key $x^r$ is defined as

\begin{equation}
\resizebox{\linewidth}{!}{$
    \text{Similarity}(x^q,x^r) =\sum_{i=1}^F\log\frac{N-N(x_i^q)+0.5}{N(x_i^q)+0.5}\cdot\mathbf{1}(x_i^q=x_i^r),
    $}
\end{equation}
where $\mathbf{1}(\cdot)$ is the indicator function, $N$ is the number of samples in the retrieved table, and $N(x_i^q)$ is the number of samples that have the value $x_i^q$ in column $i$. For numeric features, we will use quantile discretization to convert them into categorical values before retrieval. Formally, for a numeric column $T^X_{:j}$, the $k$-th quantile threshold $b_k$ is calculated as 
\begin{equation}
    b_k = Q_{\frac{k}{B}}(T^X_{:j}),
\end{equation}
where $Q$ is the empirical quantile function, $B$ is the number of categories. With the threshold, each numeric value $v$ in $T^X_{:j}$ will be converted to the $k$-th category, where $v \leq b_k$ and $v > b_{k-1}$.

Although BM25 is a non-differentiable retrieval method, it exhibits robustness to features with diverse distributions by balancing the influence of frequent and rare values through inverse frequency weighting. We leverage it as a lightweight, pre-computable retrieval mechanism that complements the downstream neural propagation module, which retains the model’s overall learning capacity. Experiments in Section \ref{sec:exp-retrieval} also demonstrate its effectiveness.

For each table in the RDB, we will retrieve and obtain the top-$K$ similar instances for each row, as shown in Figure \ref{fig:retrieval}. To integrate the retrieval results into the original RDB, we will represent them as dummy tables. As shown on the right of Figure \ref{fig:retrieval}, the dummy table has two columns: the query's id and the retrieved similar row's id. These two columns are both foreign keys that refer to the primary key of the original retrieved table. As a result, each table in the original RDB will be connected to a dummy table to incorporate the retrieval information. This approach maintains the inherent form of relational databases, allowing any existing RDB machine learning frameworks. 

\subsubsection{Graph Construction}
A common approach to modeling an RDB is to treat it as a heterogeneous graph and apply a GNN to capture and propagate the intricate dependencies and interactions. We adopt this approach to model the composite dependencies built in the retrieval module. The most intuitive way to construct a graph is by converting rows into nodes and FK-PK connections into edges \cite{cvitkovic2020supervised}, named R2N. Additionally, 4DBInfer \cite{wang4dbinfer} proposes an extension, R2N/E, which is based on R2N but specially converts the relational tables with no primary key but two foreign keys into edges. Other methods, such as hypergraphs \cite{bai2021atj}, are similar to R2N. In SRP, we try both R2N and R2N/E and choose the one with better performance on the validation set.

\subsubsection{Propagation}\label{sec:propagation}
Once the relational database is transformed into a heterogeneous graph, GNN is employed to enhance node representations by iteratively updating each node’s representation through feature aggregation from its neighboring nodes. These neighbors include both naturally connected ones through FK-PK relationships and retrieved ones through the retrieval module, incorporating both unary and composite dependencies.

During each iteration, nodes exchange information with their neighbors through a process known as message passing] \cite{gilmer2017neural}. Each node sends its current embedding to its neighbors and receives embeddings from them, which are then aggregated to capture adjacent patterns. Formally, 
\begin{equation}
    \mathbf{h}^l_v=f(\{\mathbf{h}^{l-1}_u|u\in\mathcal{N}^v \},\mathbf{h}^{l-1}_v;\theta),
\end{equation}
where $\mathbf{h}^l_v$ is the embedding of node $v$ in the $l$-th layer of GNN. $\mathcal{N}^v$ represents the set of neighbors of node $v$. $f$ is a aggregation function with parameters $\theta$ to fuse information from neighbor nodes. SRP is adaptable to various GNN architectures. To ensure a broad exploration of options, we employ widely-used architectures RGCN \cite{schlichtkrull2018modeling}, RGAT \cite{busbridge2019relational}, HGT \cite{hu2020heterogeneous}, RPNA \cite{corso2020principal}, and choose the one with the best performance on the validation set.

Combining retrieval and propagation modules, the composite dependency modeling process can be expressed as 
\begin{equation}
    \mathbf{H_c}=P(R(D);\theta),
\end{equation}
where $R$ and $P$ represent the retrieval and propagation modules, respectively. 

\subsection{Prediction}
The prediction module is aimed at calculating the final output based on the node embedding $\mathbf{H_c}$ and feature embedding $\mathbf{H_u}$, fusing the unary and composite dependencies. We concatenate these vectors to form a comprehensive representation, followed by a multilayer perceptron network to reduce dimension and learn complex patterns. The final prediction is
\begin{equation}
    \hat{y}=\sigma(\text{MLP}(\mathbf{H_u}\oplus \mathbf{H_c})),
\end{equation}
where $\hat{y}$ represents the predicted output and $\sigma$ is the activation function. For regression tasks, a linear activation function is used, whereas for classification tasks, a softmax activation function is employed to output probabilities.

\section{Experiments}

\subsection{Experimental Setup}


\subsubsection{Datasets and Metrics}

\begin{table}[htbp]
\centering
\caption{Statistics of each dataset.}
\label{tab:dataset}
{ \setlength{\tabcolsep}{1mm}
\begin{tabular}{ccccc}
\toprule
Dataset       & \# Tables & \# Columns &    \# Rows    \\
\midrule
Amazon Review (AZ)       &     3     &     15     &   16,073,957  \\
RetailRocket (RR)  &     3     &     11     &   24,885,613  \\
Outbrain (OB)      &     8     &     31     &   4,778,952 \\
StackExchange (SE) &     7     &     49     &    6,140,680  \\
Seznam Wallet (SZ)        &     4     &     14     &    2,688,678  \\
\bottomrule
\end{tabular}
}
\end{table}

We evaluate SRP on five large-scale real-world datasets from diverse domains, summarized in Table~\ref{tab:dataset}. The tasks include Churn Prediction on Amazon Book Review (AZ)~\cite{amazon-reviews} and StackExchange (SE)~\cite{stackexchange}, Conversion Rate Prediction on Retailrocket (RR)~\cite{retailrocket}, Click-Through Rate Prediction on Outbrain (OB)~\cite{outbrain-click-prediction}, and Charge Type Classification on Seznam Wallet (SZ)~\cite{motl2015ctu}. We use AUC for binary classification tasks (AZ, RR, OB, SE), and accuracy (ACC) for the multi-class task (SZ).

\begin{table*}[htbp]
    \centering
    \caption{Performance comparison. The best results are in bold, and the second-best results are underlined. ``Rel. Impr.'' means the relative improvement of SRP against the baselines. The ensemble method is not compared since it trains three models instead of one model. * marks statistically significant improvements over the sub-optimal results with $p < 0.05$ in 5 trials.}
    \label{tab:overall_result}
{\small \setlength{\tabcolsep}{1mm}
\begin{tabular}{ccccccccccccc}
\toprule
        \multirow{2}{*}{Group} & \multirow{2}{*}{Model} & \multicolumn{2}{c}{AZ} & \multicolumn{2}{c}{RR} &  \multicolumn{2}{c}{OB} & \multicolumn{2}{c}{SE} & \multicolumn{2}{c}{SZ} \\ \cline{3-12} 
 & & AUC$\uparrow$ & Rel. Impr. & AUC$\uparrow$ & Rel. Impr. & AUC$\uparrow$ & Rel. Impr. & AUC$\uparrow$ & Rel. Impr. & ACC$\uparrow$ & Rel. Impr. \\
\midrule
\multirow{5}{*}{Join} &
MLP & 0.5642 & 39.90\% & 0.5097 & 68.02\% & 0.4891 & 31.28\% & 0.6024 & 46.12\% & 0.5692 & 42.36\% \\
& DeepFM & 0.5553 & 42.14\% & 0.4933 & 73.61\% & 0.5109 & 25.68\% & 0.5984 & 47.09\% & 0.5416 & 49.61\% \\
& FTT & 0.5602 & 40.90\% & 0.4917 & 74.17\% & 0.5203 & 23.41\% & 0.6319 & 39.29\% & 0.5825 & 39.11\% \\
& XGBoost & 0.5510 & 43.25\% & 0.5000 & 71.28\% & 0.5000 & 28.42\% & 0.5820 & 51.24\% & 0.5878 & 37.85\% \\
& AG & 0.5712 & 38.18\% & 0.5096 & 68.05\% & 0.4969 & 29.22\% & 0.5820 & 51.24\% & 0.5938 & 36.46\% \\
\midrule
\multirow{5}{*}{DFS} &
MLP & 0.6815 & 15.82\% & 0.8181 & 4.68\% & 0.5456 & 17.69\% & 0.8326 & 5.72\% & 0.7554 & 7.27\% \\
& DeepFM & 0.6667 & 18.39\% & 0.8182 & 4.67\% & 0.5289 & 21.40\% & 0.8212 & 7.18\% & 0.7016 & 15.49\% \\
& FTT & 0.6765 & 16.67\% & 0.8035 & 6.58\% & 0.5360 & 19.79\% & 0.8376 & 5.09\% & 0.7473 & 8.43\% \\
& XGBoost & 0.6922 & 14.03\% & 0.7906 & 8.32\% & 0.5421 & 18.45\% & 0.8251 & 6.68\% & 0.7600 & 6.62\% \\
& AG & 0.7291 & 8.26\% & 0.8008 & 6.94\% & 0.5494 & 16.87\% & 0.8396 & 4.84\% & 0.7731 & 4.81\% \\
\midrule
\multirow{4}{*}{R2N} &
RGCN & 0.7358 & 7.27\% & 0.8470 & 1.11\% & 0.6239 & 2.92\% & 0.8558 & 2.85\% & 0.7917 & 2.35\% \\
& RGAT & 0.7410 & 6.52\% & 0.8284 & 3.38\% & 0.6146 & 4.47\% & 0.8645 & 1.82\% & 0.8026 & 0.96\% \\
& RPNA & \underline{0.7551} & 4.53\% & 0.8366 & 2.37\% & 0.6249 & 2.75\% & 0.8664 & 1.59\% & 0.8000 & 1.29\% \\
& HGT & 0.7543 & 4.64\% & \underline{0.8495} & 0.81\% & 0.6260 & 2.57\% & \underline{0.8670} & 1.52\% & 0.7965 & 1.73\% \\
\midrule
\multirow{4}{*}{R2N/E} &
RGCN & 0.7207 & 9.52\% & 0.8091 & 5.85\% & 0.6271 & 2.39\% & 0.8485 & 3.74\% & 0.7842 & 3.33\% \\
& RGAT & 0.7258 & 8.75\% & 0.7536 & 13.64\% & 0.6308 & 1.79\% & 0.8528 & 3.21\% & \underline{0.8046} & 0.71\% \\
& RPNA & 0.7348 & 7.42\% & 0.8427 & 1.63\% & 0.6322 & 1.57\% & 0.8657 & 1.67\% & 0.7988 & 1.44\% \\
& HGT & 0.7208 & 9.50\% & 0.8342 & 2.66\% & \underline{0.6323} & 1.55\% & 0.8560 & 2.83\% & 0.7983 & 1.50\% \\
\midrule
& \textit{Ensemble} & \textit{0.7228} & - & \textit{0.8522} &- & \textit{0.6290} &- & \textit{0.8678} &- & \textit{0.7912} &- \\
\midrule
& SRP & \textbf{0.7893*} & - & \textbf{0.8564*} &- & \textbf{0.6421*} & -& \textbf{0.8802*} &- & \textbf{0.8103*} &- \\
\bottomrule
\end{tabular}
}
\end{table*}

\subsubsection{Compared Methods} 
We compare SRP with four sets of baselines. The first group consists of several tabular prediction models: XGBoost \cite{chen2016xgboost}, DeepFM \cite{guo2017deepfm}, FT-Transformer (FTT) \cite{gorishniy2021revisiting}, and AutoGluon (AG) \cite{erickson2020autogluon}. These models are applied only to the target table along with information that can be easily joined from other tables.  The second uses DFS~\cite{kanter2015deep} to synthesize new features into the target table, followed by the same tabular models. The third and fourth groups transform the RDB into graphs using R2N~\cite{cvitkovic2020supervised} or R2N/E~\cite{wang4dbinfer}, and apply powerful heterogeneous GNNs, including RGCN \cite{schlichtkrull2018modeling}, RGAT \cite{busbridge2019relational}, HGT \cite{hu2020heterogeneous}, and RPNA \cite{corso2020principal}, to make predictions. Other GNN-based models mentioned in Section \ref{sec:related-work} are not included due to a lack of open-source implementations.  We also compare with an ensemble method detailed in Section \ref{sec:overall-performance}.

\subsubsection{Experiment Details}
We implement SRP and obtain all experimental results based on the RDB benchmark 4DBInfer \cite{wang4dbinfer} for fair comparisons. Hyperparameters are tuned with random search over 100 trials. The model is trained on the training set, the best hyperparameters are selected on the validation set, and the results are reported on the test set. In the synthesis, retrieval, and propagation process, we ensure there is no information leakage by filtering out future entities according to the timestamp. More experimental details can be found in Appendix \ref{app:hpps}.

\subsection{Overall Performance}\label{sec:overall-performance}

The overall performance comparison is shown in Table \ref{tab:overall_result}, from which we can have the following observations: (i) Across all datasets, the SRP framework consistently and significantly outperforms both traditional tabular models and advanced graph-based models. This is attributed to the integration of synthesis, retrieval, and propagation modules, allowing SRP to capture both the explicit FK-PK and implicit retrieved dependencies within the relational databases. (ii) Graph-based models generally outperform tabular models, which can be partly attributed to the trainable message passing mechanism that learns adjacent patterns. (iii) Although feature synthesis combined with tabular models is overall not the best approach, it still shows significant improvement over simple joins, highlighting the potential of incorporating feature synthesis to explore unary dependency.

To empirically validate the advantages of SRP as a whole compared to the ensemble models, we implemented an ensemble approach \cite{boateng2023ensemble} with the isolated synthesis, retrieval, and propagation models. Specifically, we first train these three models independently on the training set, and then obtain the prediction results of each model. Subsequently, an MLP is trained using these individual predictions as inputs to generate a final aggregated prediction. The experimental results indicate that SRP consistently outperforms the ensemble models. This improvement underscores the effectiveness of our unified SRP framework compared to isolated approaches.

\subsection{Ablation Study}




\begin{table}[htbp]
    \centering
    \caption{Ablation study on different modules. $S$, $R$, and $P$ refer to the  Synthesis, Retrieval, and Propagation modules, respectively. A.R. represents the average rank.}
    \label{tab:ablation:module}
{\small \setlength{\tabcolsep}{1mm}
\begin{tabular}{cccccccccc}
\toprule
   $S$ &    $R$ &  $P$ & AZ &  RR &  OB & SE & SZ & A.R. \\ 
\midrule
    $\times$ &     $\times$ &     $\times$ &     0.5000 &  0.5117 &  0.4861 &  0.4982 &  0.4381 & 8.0 \\
$\checkmark$ &     $\times$ &     $\times$ &  0.6783 &  0.7998 &  0.5509 &  0.8410 &  0.7537 & 6.8 \\
    $\times$ & $\checkmark$ &     $\times$ &  0.7418 &  0.7631 &  0.6158 &  0.8466 &  0.7614 & 6.2 \\
    $\times$ &     $\times$ & $\checkmark$ &  0.7551 &  0.8495 &   0.6323 &   0.8670 &  \underline{0.8046} & 3.2 \\
$\checkmark$ & $\checkmark$ &     $\times$ &  0.7713 &  0.8379 &  \underline{0.6406} &  0.8637 &  0.7961 & 4.2 \\
$\checkmark$ &     $\times$ & $\checkmark$ &  \underline{0.7834} &  0.8435 &  0.6312 &  \underline{0.8757} &  0.8037 & \underline{3.0} \\
    $\times$ & $\checkmark$ & $\checkmark$ &  0.7752 &  \underline{0.8531} &  0.6297 &  0.8662 &  0.8017 & 3.6 \\
$\checkmark$ & $\checkmark$ & $\checkmark$ &  \textbf{0.7893} &  \textbf{0.8564} &  \textbf{0.6421} &  \textbf{0.8802} &  \textbf{0.8103} & \textbf{1.0} \\
\bottomrule
\end{tabular}
}
\end{table}

\subsubsection{Modular ablation} 
Here, we evaluate the impact of each SRP module and their combinations on the overall performance, as shown in Table \ref{tab:ablation:module}.

\textbf{Synthesis Module} Compared to the retrieval and propagation modules, the synthesis module contributes relatively less when activated in isolation. However, it plays a crucial complementary role when combined with either retrieval or propagation. Notably, removing the synthesis module leads to an average rank drop to 3.6 across five datasets, underscoring its indispensability to the overall framework.

\textbf{Retrieval Module} Enabling only the retrieval module yields notable performance gains, indicating that retrieving similar samples introduces a useful inductive bias for modeling composite dependencies. Incorporating the retrieval module consistently enhances the performance of the SRP framework across datasets. Notably, the combination of the retrieval and synthesis module achieves the second-best result on the OB dataset. This finding challenges the prevailing paradigm in which GNN-based methods dominate RDB prediction, highlighting a promising alternative that does not rely on graph neural networks.

\textbf{Propagation Module} Graph neural network is already a strong baseline, which can achieve relatively good performance itself. However, simply using the propagation module cannot achieve the optimal performance. Combining the synthesis or retrieval module with it can further boost the predictive accuracy.

\subsubsection{Dataset Level Analysis}
From Table~\ref{tab:ablation:module}, we observe that certain module combinations can sometimes achieve relatively strong results. For example, the Churn Prediction tasks on AZ and SE are defined on the \verb|User| table, which contains limited user attributes due to privacy constraints. As a result, the retrieval module may not be able to bring much useful information by retrieving similar samples. 
In Retailrocket, where only 2.5\% of the samples are positive, the class imbalance presents a significant modeling challenge. Although using all three SRP modules results in a high validation AUC of 0.8632, the performance is marginally better than that of the variant without the synthesis module. Since both the synthesis and propagation modules are capable of modeling unary dependencies along FK-PKs, there is a risk of overfitting on the validation set for imbalanced datasets.
For the Outbrain dataset, combining the synthesis and retrieval modules achieves strong performance. One possible explanation is that Outbrain is a sparse dataset, where most users have few interactions. In such cases, the propagation module may contribute little and can even be unnecessary.

Despite such variations, the full SRP model consistently achieves the best overall performance, demonstrating the complementary strengths of its three modules in capturing complex relational dependencies.


\subsubsection{Dense Retrieval} \label{sec:exp-retrieval}
In addition to using sparse retrieval methods (BM25 \cite{robertson1995okapi}) to measure the similarity in the retrieval module, we also tried dense retrieval methods \cite{zheng2023dense} to find a better measurement. The results are shown in Table \ref{tab:dense}. We can see that sparse retrieval shows more consistent and superior performance. The mainstream retrieval method for tabular data is still sparse retrieval. Since tabular data is highly structured and discrete, dense retrieval may not be as effective as sparse retrieval.
\begin{table}[htbp]
    \centering
    \caption{Comparison of sparse and dense retrieval.}
    \label{tab:dense}
    {\small \setlength{\tabcolsep}{1mm}
    \begin{tabular}{cccccc}
    \toprule
      & AZ & RR & OB & SE & SZ \\
    \midrule
    Dense & 0.7807 & 0.8358 & 0.6313 & 0.8675 & 0.7870 \\
    Sparse & \textbf{0.7893} & \textbf{0.8564} & \textbf{0.6421} & \textbf{0.8802} & \textbf{0.8103} \\
    \bottomrule
    \end{tabular}
    }
\end{table}

\subsubsection{Frequency-aware Aggregator}

Table \ref{tab:fa-aggregator} demonstrates the effectiveness of our proposed frequency-aware (\verb|FA|) aggregator in the synthesis module. To balance performance and efficiency, we set the reserved categories to $m=2$. From the result, we can find that with richer categorical information obtained from the \verb|FA| aggregator, the synthesis module can more effectively model unary dependencies, leading to improved performance.
\begin{table}[htbp]
    \centering
    \caption{The influence of the frequency-aware aggregator.}
    \label{tab:fa-aggregator}
    {\small \setlength{\tabcolsep}{1mm}
    \begin{tabular}{cccccc}
    \toprule
      & AZ & RR & OB & SE & SZ \\
    \midrule
    w/o \texttt{FA} & 0.7831 & 0.8520 & 0.6377 & 0.8751 & 0.8067 \\
    w/ \texttt{FA} & \textbf{0.7893} & \textbf{0.8564} & \textbf{0.6421} & \textbf{0.8802} & \textbf{0.8103} \\
    \bottomrule
    \end{tabular}
    }
\end{table}

\subsection{Hyperparameter Study}\label{sec:hpo}

\begin{figure}[htbp]
    \centering
    \includegraphics[width=1\linewidth]{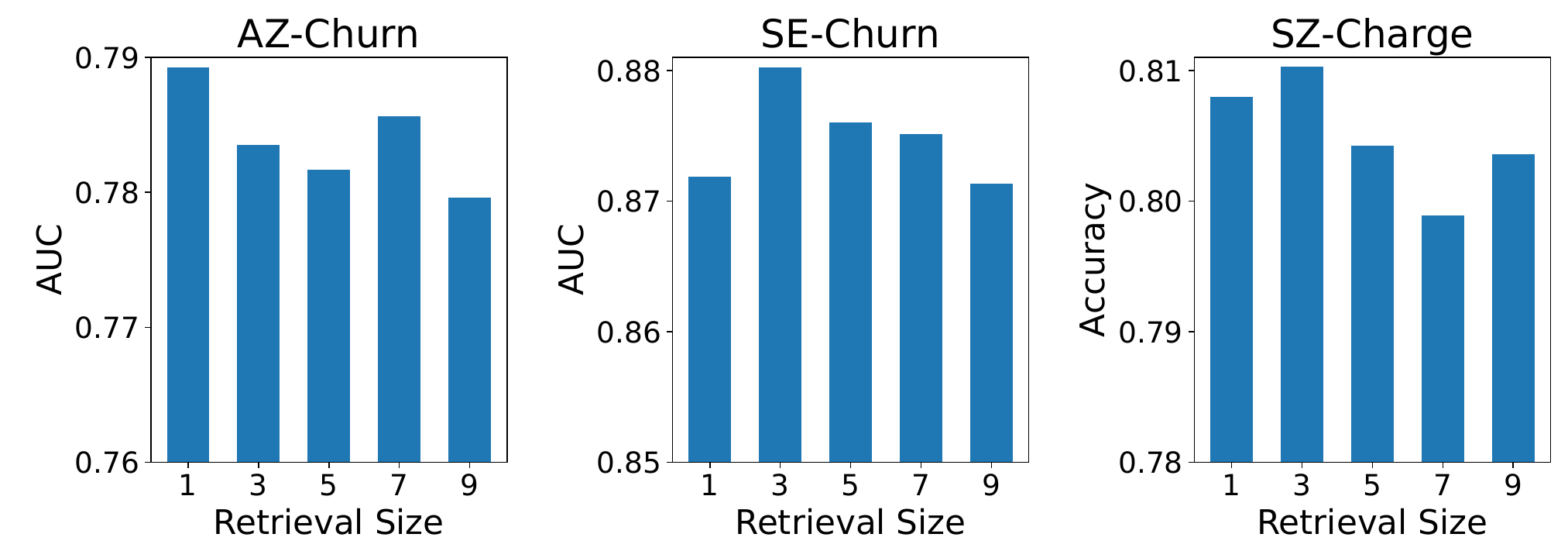}
    \caption{Performance of SRP w.r.t. different retrieval sizes.}
    \label{fig:different-k}
\end{figure}

In the retrieval module, we select the top-$K$ most similar rows based on the computed similarity scores. Figure \ref{fig:different-k} illustrates the performance of SRP under varying retrieval sizes $K$ across datasets of different scales. The results indicate that a small retrieval size can lead to strong performance. As $K$ increases, the inclusion of less relevant or noisy entries may degrade overall performance. While performance may fluctuate slightly with different values of $K$, SRP generally performs well across a broad range. In practice, we recommend selecting a relatively small value of $K$ to balance performance and efficiency. The results reported in Table \ref{tab:overall_result} are chosen from the better outcome between $K=1$ and $K=3$.

\subsection{Efficiency and Scalability}
To ensure practical applicability in real-world scenarios such as fraud detection and advertising, SRP is designed with efficiency and scalability in mind. As shown in Figure \ref{fig:framework}, the feature synthesis and retrieval processes are non-parametric, allowing them to be pre-computed offline. The time complexity analysis of SRP is provided in Appendix \ref{app:time-complexity}. Additionally, we compared the average inference time of SRP and RGCN, as shown in Table \ref{tab:inference-time}. While SRP may introduce a slight increase in computational overhead (less than 10\%), this trade-off is justified by its enhanced predictive power and flexibility in capturing complex dependencies.

As for the scalability, despite operating on datasets with over ten million rows, all experiments were conducted on a single machine with 16 GB GPU and 64 GB CPU memory. For extremely large and complex RDBs, one key advantage of SRP is its modularity. Each module in SRP is pluggable, allowing users to remove certain modules to create a more lightweight solution tailored for large-scale RDBs.
\begin{table}[htbp]
    \centering
    \caption{Average inference times of SRP and RGCN.}
    \label{tab:inference-time}
    {\small \setlength{\tabcolsep}{1mm}
    \begin{tabular}{cccccc}
    \toprule
      & AZ & RR & OB & SE & SZ \\
    \midrule
    SRP ($\mu s$) & 72.91 & 235.06 & 154.52 & 97.97 & 89.23 \\
    RGCN ($\mu s$) & 66.85 & 231.10 & 141.43 & 100.12 & 81.02 \\
    \bottomrule
    \end{tabular}
    }
\end{table}

\section{Conclusion}

This paper introduced SRP, a unified framework for predictive modeling on relational databases, leveraging both unary and composite dependencies. To the best of our knowledge, SRP is the first to incorporate a retrieval mechanism into predictive tasks for RDB (multiple tables). Extensive experiments show that SRP consistently achieves state-of-the-art performance. Future work will focus on further modeling the interactions between different types of dependencies.

\bibliography{aaai2026}

\appendix

\section{Time Complexity}\label{app:time-complexity}
Consider an RDB $D$ with $N$ tables, where each table $T^i\in D$ has $R_i$ rows and $C_i$ columns. The target table is $T^t$. Here we only consider the trainable components in SRP. In the unary dependency modeling process, assume that there are $F$ features, including the original features and the generated features, to be embedded in $d_u$ dimensions. The time complexity of feature embedding is $\mathcal{O}(Fd_u^2)$. In the propagation process, suppose that the maximum number of neighbors sampled for each node is $M$. If the node embedding is $d_c$ dimension, the total time complexity to update a target node is $\mathcal{O}(C_tM(d_c^2+Md_c))$. Finally, using an MLP with $L_{fc}$ layers and $d_p$ hidden dimension, the prediction module will cost $\mathcal{O}(L_{fc}d_p^2)$. In practice, $F$ is usually is proportional to $NC_t$, and $M$ and $L$ are constants. If we represent all embedding dimension as $d$, the overall time complicity can be represented as $\mathcal{O}(NC_td^2)$.

\section{Implementation Details}\label{app:hpps}
We use random search to determine the hyperparameters. The range of each value is shown in Table \ref{tab:app-hyperparameters}. The implementation and the final hyperparameters of SRP are available in the source code.

\begin{table}[H]
    \centering
    \caption{Hyperparameter grid.}
    \label{tab:app-hyperparameters}
    {\small \setlength{\tabcolsep}{1mm}
    \begin{tabular}{cc}
    \toprule
        Hyperparameter & Values \\
         \midrule
         Learning rate & [$10^{-4}$, $10^{-2}$] \\
         Batch size & [128, 4096]\\
         Embedding size & [8, 256]\\
         Hidden size & [16, 256]\\
         Dropout & [0, 1]\\
         Number of layers & [1, 8]\\
         Attention heads  & [1, 8]\\
         GNN layers & \{1, 2, 3\}\\
         Neighbor sampling fanout & \{1, 5, 10, 20\}\\
         Number of retrieved neighbors & [1, 10]\\
         PNA aggregators & {Mean, Min, Max}\\
         Graph contruction methods & {R2N, R2N/E} \\
         SRP propagation model & {RGCN, RGAT, RPNA, HGT} \\
         \bottomrule
    \end{tabular}
    }
\end{table}



\end{document}